\documentclass[conference]{IEEEtran}
\IEEEoverridecommandlockouts

\usepackage{times}

\usepackage{graphicx}

\def\BibTeX{{\rm B\kern-.05em{\sc i\kern-.025em b}\kern-.08em
    T\kern-.1667em\lower.7ex\hbox{E}\kern-.125emX}}

\graphicspath{ {images/} }

\begin{document}

\title{5G for the Factory of the Future: Wireless Communication in an Industrial Environment}

\author{\IEEEauthorblockN{
Florian Voigtl\"ander\IEEEauthorrefmark{1},
Ali Ramadan\IEEEauthorrefmark{2},
Joseph Eichinger\IEEEauthorrefmark{2}, 
J\"urgen Grotepass\IEEEauthorrefmark{2}, 
Karthikeyan Ganesan\IEEEauthorrefmark{2}, \\
Federico Diez Canseco\IEEEauthorrefmark{3},
Dirk Pensky\IEEEauthorrefmark{4} and
Alois Knoll\IEEEauthorrefmark{1}}

\IEEEauthorblockA{\IEEEauthorrefmark{1}Technische Universit\"at M\"unchen TUM; M\"unchen, Germany; florian.voigtlaender@tum.de, knoll@in.tum.de}
\IEEEauthorblockA{\IEEEauthorrefmark{2}Huawei German Research Center GmbH; M\"unchen, Germany;\\ (ali.ramadan, joseph.eichinger, juergen.grotepass, karthikeyan.ganesan)@huawei.com}
\IEEEauthorblockA{\IEEEauthorrefmark{3}Technologie-Initiative SmartFactory-KL e.V.; Kaiserslautern, Germany;
diezcanseco@smartfactory.de}
\IEEEauthorblockA{\IEEEauthorrefmark{4}Festo Didactic SE; Denkendorf, Germany; dirk.pensky@festo.com}
}


\maketitle


\begin{abstract}

In this paper, the application of 5G communication technology in an industrial environment is discussed. It acts as an enabler for the separation of sensors/actors and resources, like memory and computational power. 5G offers characteristics essential for the proposed approach like robustness, ultra-low latency, high data rates and massive number of devices. 

A demonstrator of a production line was used as an test environment for 5G in a real-world industrial application. A wide variety of heterogeneous sensor systems is used by a mobile robot platform. The collected data is transmitted via a 5G network to various Cloud systems. The product is treated as a cyber-physical system with a RFID tag in conjunction with the product memory system. 

The dynamic production flow approach is discussed centered around the robot which is used for transportation and inspection of products. This inspection is performed during the transportation and influences the production flow directly. This is desirable in the scope of Industry 4.0 to have an efficient production down to batch size 1. 
 
\end{abstract}


\IEEEpeerreviewmaketitle

\section{Introduction} 

Industry 4.0 is a paradigm to enhance the productivity and economic performance of modern industry. 
After the introduction of computer numerical controls, it is seen as the next major leap.
A core motivation is to efficiently produce goods down to batch size 1. 
This means, that the high individuality of a product must be compensated by the flexibility of production line(s). 
In the scope of this paper, a flexible production flow enabled by a mobile robot platform is introduced in an industrial environment. 
A central problem of mobile robots is, that they require a lot of computational power and other resources to fulfill their function. A separation of physical interfaces like actors and sensors from resources like computational power and memory is desirable. 

\begin{figure}[h] 
	\centering
	{\includegraphics[width=0.75\columnwidth]{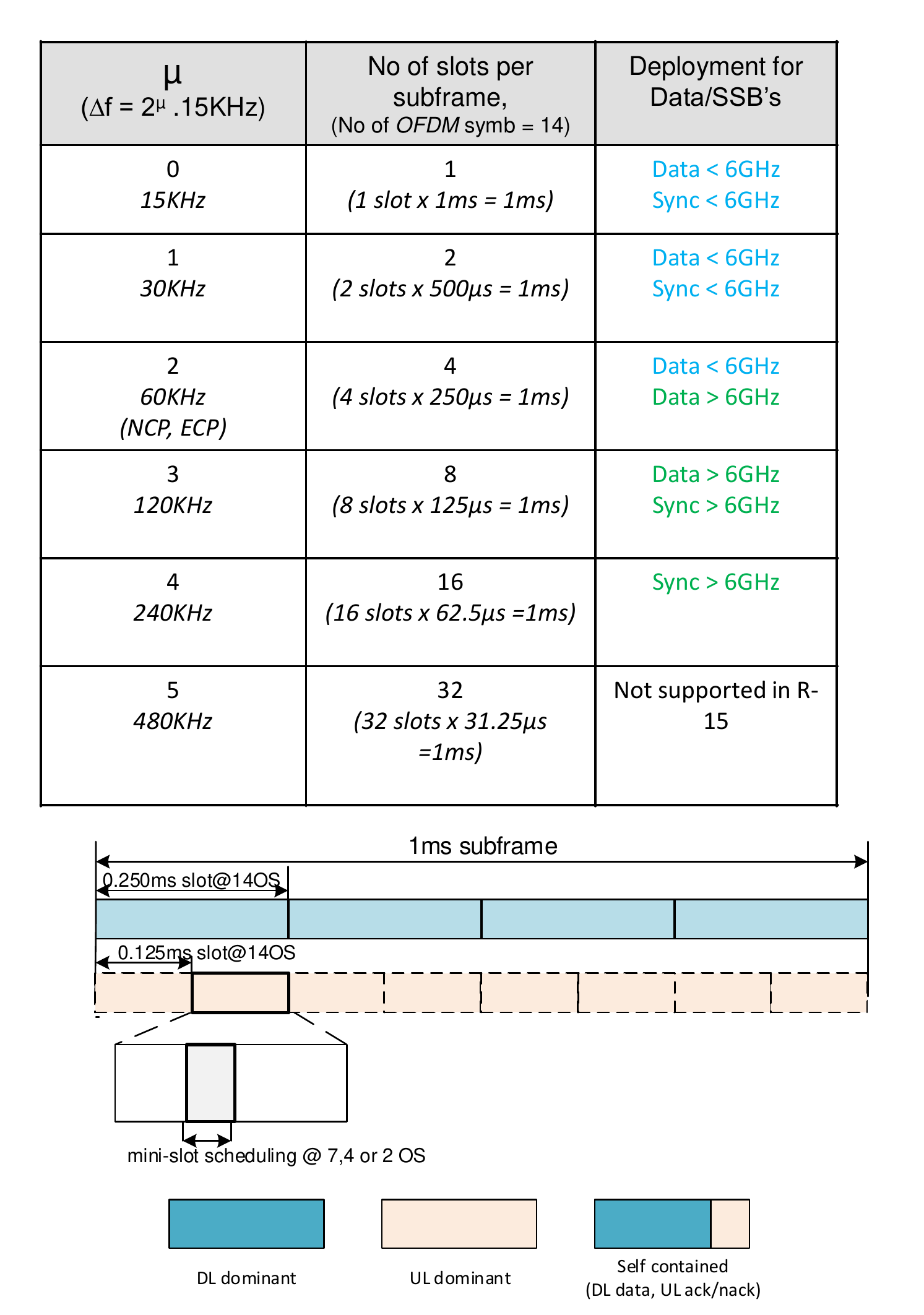}}
	\caption{5G frame structure with multiple numerology and flexible deployments}
	\label{fig:5gnrfs}
\end{figure}

The key technology for the proposed approach is the fifth generation wireless systems networking standard, commonly abbreviated as 5G NR (New Radio). NR is expected to offer a superior performance to 4G in all regards including higher data rates providing enhanced mobile broadband(eMBB), ultra-low latency and higher reliability (URLLC) \cite{4g3g}. NR achieves lower latency with a round-trip time of 1ms~\cite{fw5g}. The 5G NR frame structure with different scheduling interval is shown in figure~\ref{fig:5gnrfs}. 

The 5G radio frame is designed for 10ms duration with 1ms subframe period but NR provides the opportunity for slot based scheduling where each slot contains 14 Orthogonal Frequency Division Modulation (OFDM) symbols. NR also supports non-slot based scheduling option in the form of mini-slot with 7, 4 or 2 OFDM symbols to support ultra-low latency application. NR design features a self-contained frame structure, thereby 14 OFDM symbols in a slot could be classified as Downlink (DL), Flexible (X) and Uplink (UL) with the help of slot format indicator which provides an index to the preconfigured table.

\begin{figure}[h] 
	\centering
	{\includegraphics[width=0.75\columnwidth]{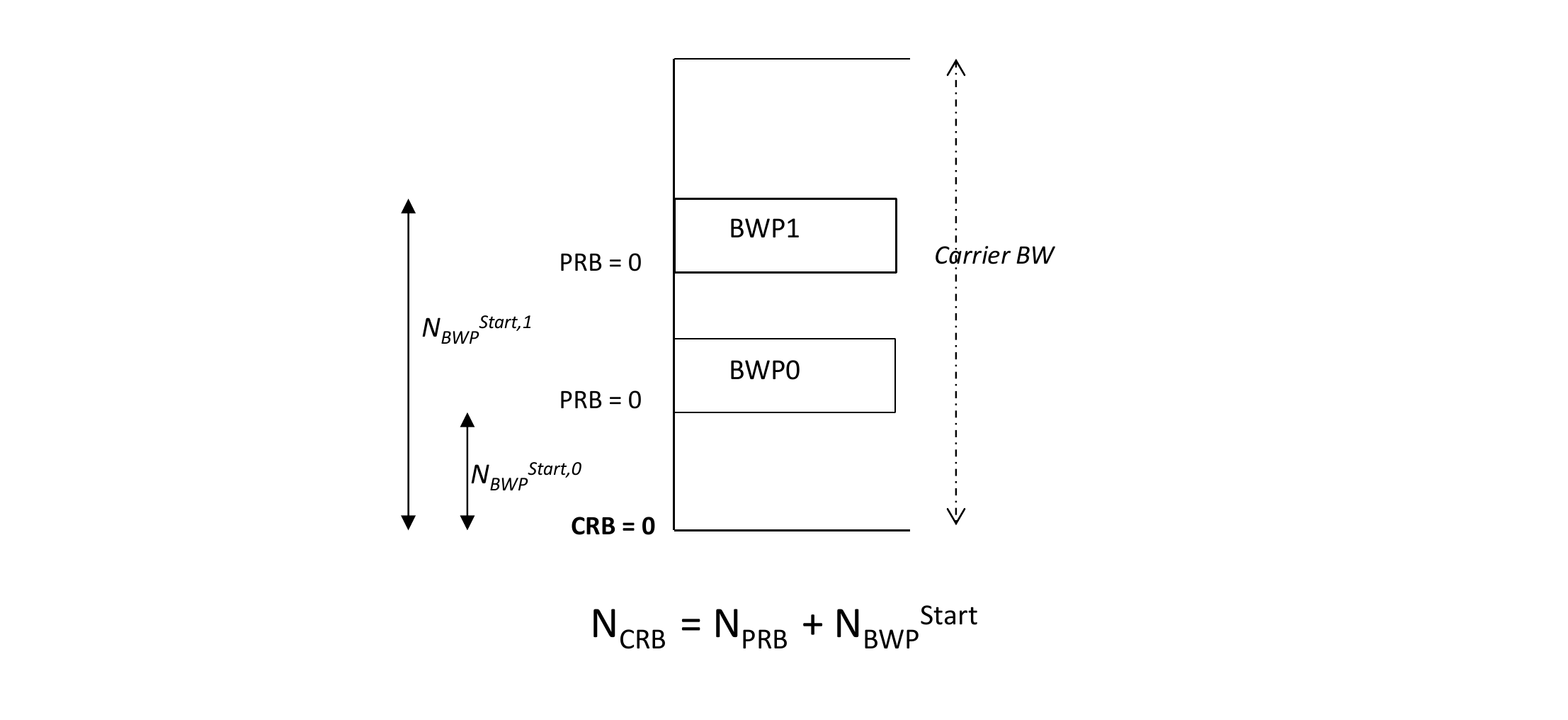}}
	\caption{Flexible bandwidth partitioning with P-OFDM}
	\label{fig:bwp}
\end{figure}

Another NR feature is the bandwidth part (BWP) which consists of a group of contiguous physical resource blocks (PRB), where each BWP configuration includes subcarrier spacing (SCS), cyclic prefix (CP), control resource set (CORESET), bandwidth size and frequency location as shown in figure~\ref{fig:bwp}. With advanced waveform technology like polar OFDM (P-OFDM), Filtered OFDM (F-OFDM) etc., a larger carrier bandwidth is flexibly partitioned to different numerology that can satisfy different applications like eMBB and URLLC. 

\section{System Requirements}

Having focused on robustness and low latency in a previous work~\cite{5gforrobotics}, 
the exploration high data volume and heterogeneous sensor systems posed an interesting topic.

In general, robotic systems use multiple different sensors to perceive their environment which generates a heterogeneous data stream. For each stream, different requirements apply, like response times, data rate and robustness. One stream may be used for a distributed control, whereas another may be used for video streaming. 
Another source of heterogeneous data is the communication with multiple systems, like different components of a production line and other mobile robots. The key requirements for each data stream is defined by the application. 
 
With real-time ultra-low latency communication, time-critical and exhaustive operations can be outsourced through a 5G network to the Cloud. 
To separate sensors and actors from the limitations of a system, a wireless communication is required for mobile application. This separation provides many benefits, like better scalability and a certain degree of independence from resources like energy, memory and computational power on a local level.

Additional to the strict requirements of the wireless communication, multiple system requirements are present. 
In an industrial environment norms and regulations need to be fulfilled, because safety is significant concern in applications. A multilayered approach is proposed, combining local safety routines with network communication.

\section{System Integration}

A demonstrator has been developed which includes multiple production line segments and a mobile robot platform which is capable of transporting products between the segments. The robot is transporting the product over the distances between the segments. Both the robot and the production line are integrated into a network with multiple Cloud systems.

\subsection{Robot Platform}
\label{subs:robotplatform}
\begin{figure}[h]
	\centering
	{\includegraphics[width=0.45\columnwidth]{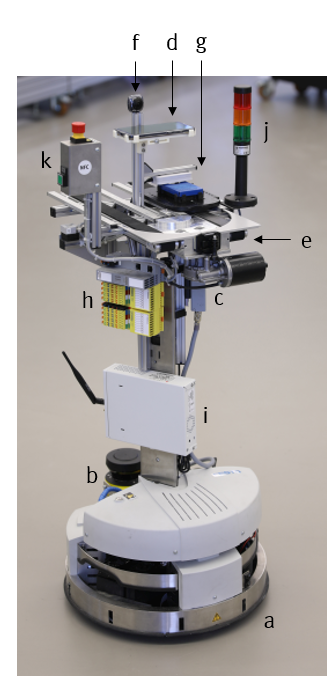}}
	\caption{The mobile robot platform with a wide variety of heterogeneous sensors: 
	a) Robotino including bumper, inertial, odometry and infrared distance sensors 
	b) Laser range sensor 
	c) converyor belt system with light barrier 
	d) smartphone used as a camera	
	e) webcam	
	f) 360 camera 
	g) RFID-reader
	h) Emergency Stop (Safety PLC)
	i) 5G radio hardware and compact computer
	j) signal column
	k) NFC-reader	
	~\cite{smfa}
	}
	\label{fig:robotandsensors}
\end{figure}

To complement the dynamic production flow, a mobile robot platform with 5G was introduced. The mobile robot possesses a holonomic drive system with three motors. A multitude of heterogeneous sensors has been attached to the robot platform as can be seen in figure~\ref{fig:robotandsensors}. 
	
A ring at the bottom portion of the platform holds 9 infrared distance sensors and a bumper which reacts to direct physical contact. The recognition of the environment has multiple layers. On long range, a laser range sensor is used. On a medium range, the infrared distance sensor array consisting of nine individual sensors is used. Additional to the surrounding environment, the docking stations are also recognized by the laser range scanner. 
	
The odometry is calculated in an dedicated FPGA which directly reads the rotary encoders of the drive motors. The hardware odometry is then used in a sensor fusion to improve the odometry based on laser range sensor and infrared distance sensors. Another on-board sensor measures the battery voltage which is also uploaded to the Cloud as an indicator for the remaining battery life. 

A compact computer performs the calculations for radio operation. This is only required in the prototyping state. Equally, the base station connecting the robot to the network also has a separate computer for radio calculations. 
The computer on the robot and the compact computer both run Ubuntu with ROS~\cite{quigley2009ros}. ROS is used as the framework for the node-based communication. An Ethernet connection allows the communication with the 5G radio prototype. A kernel patch has been performed to use a real-time linux kernel. The ROS framework requires a direct IP space visibility between devices~\cite{rosnet}. This has been realized by using a VPN tunnel~\cite{peervpn} between the robot and the computation unit. 

The mobile robot platform is equipped with multiple sensor and actor devices. A forward facing camera is used for orientation. A 360$^{\circ}$ camera generates a constant, high data volume stream which is transmitted over 5G to a distant location for a telepresence demonstration. 
Another camera is used to take pictures of the product and perform optical quality control. 
A smartphone has been used as camera to show the interoperability between widely different devices in the factory, ranging from PLCs over computers to even smart phones running Android. It is connected over USB to the Robot. The DroidCam app~\cite{droidcam} in conjunction with a linux program was used to interface with the smartphone. Apart from being used as a camera, that the smartphone provides a secondary functionality by showing process and product data live on its display. The image is uploaded into the SAP hana Cloud where trained neural networks analyze the state of the assembled components. This data is then used to influence the production flow dynamically. 

On top of the robot platform, a table with a conveyor belt is mounted. The conveyor belt is driven by a geared motor and transports the product on its tray.  
A RFID-reader scans the chip embedded into the product which holds the digital product memory. The product memory is used to compare the product with its digital twin in the Cloud and assembly plan. 
The capturing of images is triggered by the robot by recognizing the presence of a RFID tag on its conveyor belt.

\subsection{Industrial Demonstrator Platform} 

In the scope of the \textit{SmartFactory}\textsuperscript{KL} technologie initiative, a worldwide unique Industry 4.0 pilot plant was developed and presented at the Hannover Trade Fair (Hannover Messe) in 2014.~\cite{sfwp1}\cite{sfwp2}

Consisting originally of five production modules, it was later expanded to ten. It is a flexible production line that produces business card holders in a custom configuration. Each module performs only one task, as an explicit way of highlighting its modularity and the Industry 4.0 concepts. This also enables the replacement of any module (with an equivalent one or a Manual Work Station) without affecting the whole plant operation. Additionally, it provides product individuality characteristics such as batch size 1 efficient production.
	
\begin{figure}[h] 
	\centering
	{\includegraphics[width=1\columnwidth]{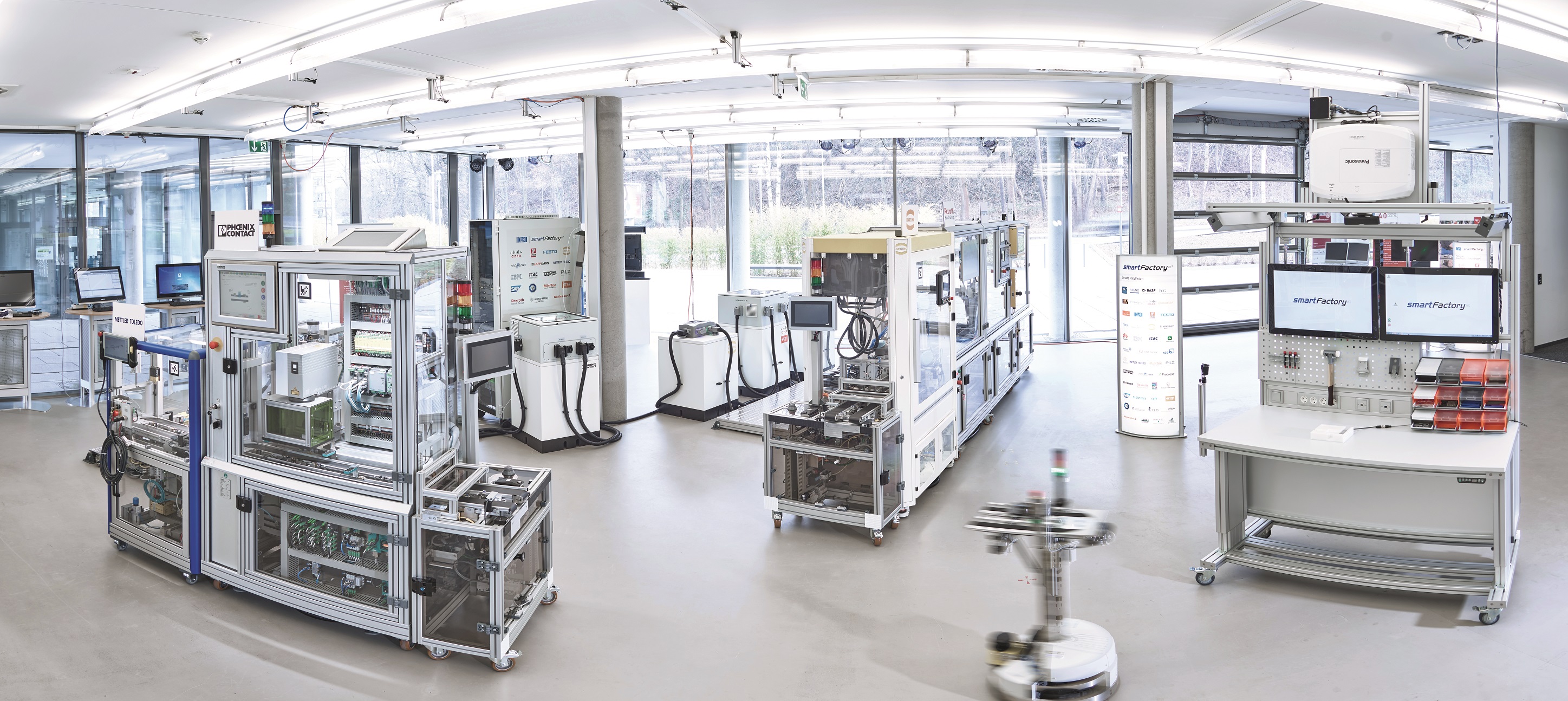}}
	\caption{Industry 4.0 demonstrator segmented into three production islands at Hannover Fair 2017~\cite{smfa}}	
	\label{fig:plant2017}
\end{figure}

The final product will consist of an engraved acrylic plate with an RFID tag. A spring is inserted into a recess in the plate and a metal cover is mounted on top of it. A laser engraves a QR code with the customer data given during the order creation. Quality controls are performed at the end in two modules, a scale and an optical inspection module. A human-operated workstation is used for extra features not covered by the standard configuration of the plant, such as extra cover colors, giveaways being inserted into the product or completely replacing a module being taken out of production for maintenance. 
As a cyber-physical system it also is connected to the network via the RFID-reading/writing units.
The data on the tag is updated according to the product's progress, so it bears its own digital twin.

For the 2017 edition, the plant was divided into two independent lines served by a Flexible Transport System (FTS), the mobile robot platform, which transports the carriers while the product is being processed between docking stations (figure~\ref{fig:plant2017}) attached at the end of each line. Alternatively it is brought to a manual work station where an operator could fulfill additional tasks not supported by the plant, or replace a module taken out from production.

The current configuration expanded on this principle as it presented three production islands forming one production line, each of which could fulfill multiple steps of the process. This reduces bottlenecks in the production process by using redundancy to balance the workload in a flexible configuration of production islands. 

The different modules are not directly connected. Each one has a conveyor belt system with a gate at each side, which only open when there is a suitable module an neighbor. A docking station enables product carriers to leave each production island by the means of the robot platform. 

The transfer of the product between modules is regulated by a central handshake controller, which reads the state of OPC-UA servers in all the modules and docking stations. The modules of the production line can be rearranged freely in space or exchanged with a different production island. Products will eventually find the way into their next production step, being transfered to a different island by the robot, if required. The physical interface between robot and a docking station is a special conveyor belt system mounted on a table on top of the robot. 

\begin{figure}[h] 
	\centering
	{\includegraphics[width=1\columnwidth]{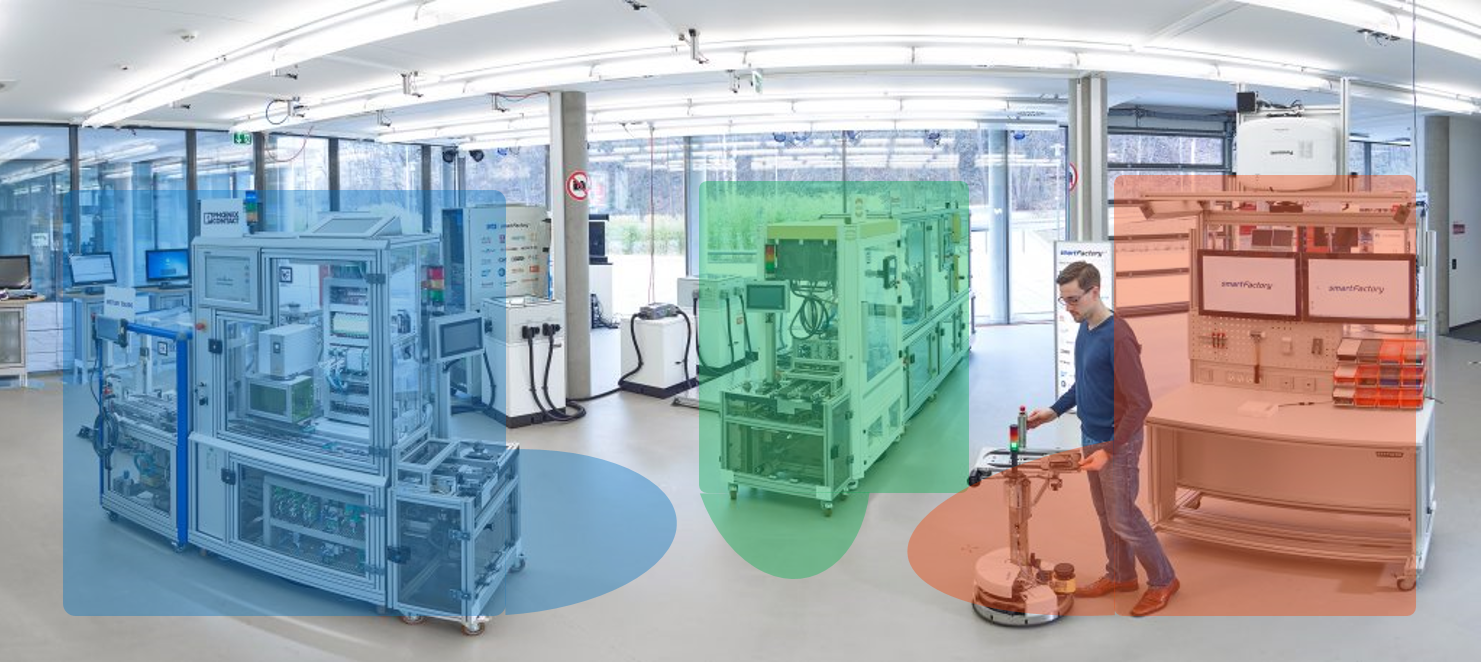}}
	\caption{Industry 4.0 Profisafe compliant robot, Hannover Fair 2018~\cite{smfa}}
	\label{fig:plant2018}
\end{figure}

The separation of the plant into three islands also involved a mayor change in the safety concept. The previous plant configuration included a hardwired redundant safety system in which the safety devices of any module would stop the whole plant. Keeping in mind the vision of a large modular production line with many islands, this behavior was not desirable. The operation of the safety devices should be confined to each island itself. 

For this, a Safety Model through Profisafe was implemented, with a central Safety PLC and Bus Couplers in the modules which could not directly manage the Profisafe protocol. The Robot, as a flexible transport system, was not included in the Safety model and operated completely independent from it.

In the context of this paper, a third docking station was implemented for the 2018 Hannover Messe, which extended the plant into 3 production islands and the Manual Work Station (figure~\ref{fig:plant2018}). A Profisafe Bus Coupler was installed into the Robot, making use of the high data rates and ultra-low latency inherent to 5G for the wireless communication of the Bus Coupler and the whole robot. 

The Robot inserts itself into the Safety Loop of the Production Island to which it is docking and acts as another module in it. It signals its affiliation to the island with colored lights matching the paneling of each docking station. After undocking and during transit, the safety behavior of the robot is isolated from that of the production islands.

Additional systems were installed on the robot as non-time-dependent communication devices, providing quality control during the transport between production islands. These included an RFID reader and a camera for optical control. The information was sent to a Cloud for further processing and decision as mentioned in  \ref{subs:robotplatform}. 

The route of the product is dynamically influenced by its state. If a mounting error occurred or the assembly station is busy, the route is diverted to a manual work station. Therefore, the production flow is highly adaptive to changes during the runtime. The quality check is performed in an otherwise dead time in the process. 
A delay caused by an unstable connection in not very critical from a safety-centered view although this delay has a direct influence on the productivity and effectiveness of the production flow. Local safety functions are implemented on the robot itself to avoid unwanted behavior.

\subsection{5G Radio Interface} 

\begin{figure}[h] 
	\centering
	{\includegraphics[width=1\columnwidth]{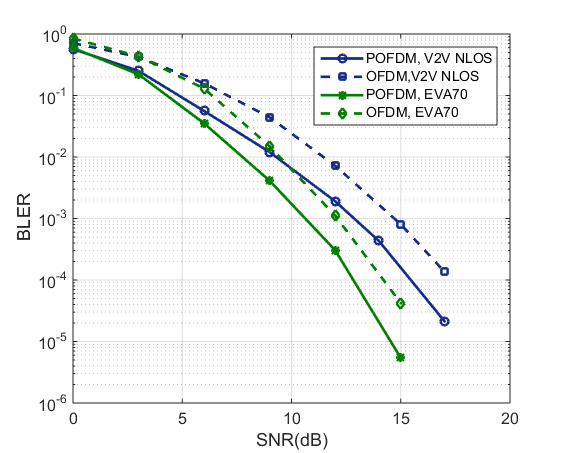}}
	\caption{Block Error Rate (BLER) over signal-to-noise ratio (SNR) performance for two different waveforms and two different channels}
	\label{fig:BLER}
\end{figure}

\begin{figure}[h] 
	\centering
	{\includegraphics[width=1\columnwidth]{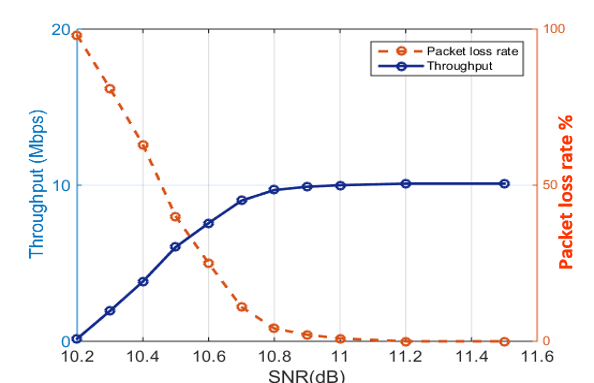}}
	\caption{P-OFDM iPerf throughput measurement with UDP traffic}
	\label{fig:Thro}
\end{figure}

The communication over 5G consists of the antenna prototypes and computers needed for radio calculations. The physical layer of the 5G system is an software-defined radio based system with highly reconfigurable system parameters. 
Both 5G UE and 5G base station (gNB) devices have two main components: Baseband unit for PHY and RF unit connected to baseband unit with gigabit ethernet. The system supports multiple carrier frequencies, e.g.  800MHz, 2.6GHz, 3.5GHz, and different bandwidth options: 5 / 10 / 20 MHz. The frame structure has a flexible TTI length: 125us, 250 us, 500us, 1ms enabling scalable latency and throughput. The basic functionalities of the physical downlink shared channel (PDSCH) are similar to that of LTE. However extra 5G based components are added to achieve high reliability and configurability to the system. Unlike LTE, both downlink and uplink channels use OFDM based waveform. To achieve different level of reliability and spectral efficiency, a poly-phase network (PPN) module is used to shape the OFDM signal to different waveform types such as P-OFDM, W-OFDM, and CP-OFDM.

Figure~\ref{fig:BLER} illustrates the physical layer performance of the 5G system. Propsim F32 channel emulator was used for the measurements. Two system configuration has been examined with two different waveform configurations; OFDM (the conventional waveform for LTE) and pulse shaped OFDM P-OFDM. From the results, P-OFDM outperforms OFDM with ~1.7 dB for both channels (3GPP EVA70 and vehicle-2-vehicle V2V Urban NLOS channel. For P-OFDM, the 99.999\% reliability, equivalent to 99.99999\% availability in case of survival time of 2 slots (requirement for I4.0), is achieved at 15 dB signal-to-noise ratio (SNR) in EVA70, and at 19 dB in V2V NLOS. System throughput is illustrated in Figure~\ref{fig:Thro} for different SNR values. The system reaches 10 Mbps at 11 dB SNR.

\section{Testing}


Multiple measurements have been performed to evaluate the performance and behavior of the prototype system. 
The distribution has been captured with the Wireshark~\cite{ws224} tool. The measurements show, that a data rate of data rate of 5.97Mbit/s was present during operation. This includes all of the packets and fits into the 5G frame structure as shown in figure \ref{fig:5gnrfs}.

\begin{figure*}[h]
	\centering
	{\includegraphics[width=\textwidth]{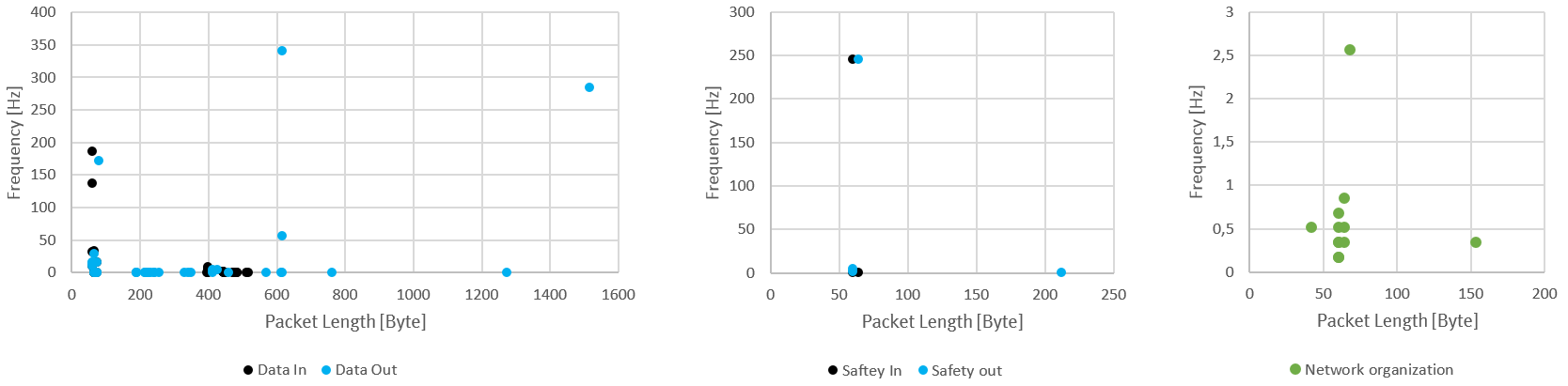}}
	\caption{The distribution of messages between the respective systems. }
	\label{fig:dist_mess}
\end{figure*}

The real-life distribution of transmitted packets have been analyzed while the system was running as can be seen in figure \ref{fig:dist_mess}.The measurements are sorted in relation to the components of the system. The packets have been categorized into three classes: non-safety relevant, safety-relevant and network organization. 
Overall, a wide variety of communication characteristics can be examined. 

Dominant packages with high frequency are the streams of the cameras. Much smaller and less frequent is the safety-relevant communication. Strict requirements are imposed on this communication. A detailed list of the safety-relevant data can be found at table \ref{tab:safe}. The majority of the communication is symmetrical between the two nodes. Packets of Length 60 and 64 are exchanged at a frequency of roughly 250Hz.
Messages with organizational functions pose only a fraction of the experienced traffic.

\begin{table}
  \centering
  \begin{tabular}{|l|l|l|r|r|}
        Source & Destination & Protocol & Len.[Byte] &  Freq.[Hz] \\
    \hline
    Hilscher   & PhoenixC & PNIO     & 60   & 246.19 \\
    Hilscher	 & PN-MC    & PN-DCP  & 60   & 0.51    \\
    PhoenixC & PN-MC    & PN-DCP  & 60   & 1.36    \\
    PhoenixC & Hilscher   & PNIO     & 64   & 246.19 \\
    PhoenixC & LLDP MC & LLDP      & 212 & 0.17    \\
    PhoenixC & LLDP MC & PN-PTCP & 60   & 4.94    \\  
  \end{tabular}
  \vspace{3mm}
  \caption{Measurements of safety-relevant ProfiSafe messages.} 
  \label{tab:safe}
\end{table}

\section{Discussion} 
	
In~\cite{5gforrobotics}, a mobile robot platform performing a ultra-reliable, low latency distributed control has been introduced. That application had only a small set of sensors which were used sending small data packages rapidly.
The proposed approach expands the scope to a multitude of heterogeneous data sources and introduces ProfiSafe messages transmitted over 5G and data transmission to cloud systems. A higher level of capability is achieved using a robot platform with a multitude of sensors in an industrial environment. Additionally, a distributed multi-layered safety approach has been introduced, enabling a dynamic production flow. 

Measurements have been taken in a real-world industrial environment to analyze and optimize 5G communication. 
Another analysis and characterization was performed on the network traffic regarding the application-specific requirements. 

The multitude of sensors allow a great variety of applications, not implemented yet. Software may provide greater functionality in the future on the same hardware platform. The 360 degrees video stream may be used for remote operation of the robot or telepresence applications as an example. 
This enables the separation of perception and action from computation on robot platforms. A central solution provides easier scalability and saves local resources on the robot, like computational power, memory and energy. 

\subsection{Comparison to 3GPP 5G standard use cases}
\begin{table}
  \centering
  \begin{tabular}{l|l|l}
     Use case aspects   		& 1 					 & 2 						\\
     \hline
	Availability                  & 99,9999\% to            & 99,9999\% to              \\        
	                                & max. $99,999999$\% & max. $99,999999$\%	\\
	\hline
	End-to-end latency: & &\\ 
	Target value	          & $<12ms$                  & $<30ms$				\\
	Jitter                         & $<6ms$ 	                 & $<15ms$				\\
	\hline
	Service data rate		&						 & $>5Mbit/s$ 				\\
	\hline
	Message size	 	     & $40$ to $250Byte$     &							\\
	\hline
	Transfer interval          & $12ms$                    &							\\
	\hline
	Survival Time              & $12ms$                    &							\\
	\hline
	Service area	                & max. $200m$ x $300m$ &						\\	
    \hline 
  \end{tabular}
  \vspace{3mm}
  
  \caption{ 
  Excerpt from~\cite{3gpp_rel16}, use case 5.3.6. with different aspects. 
  } 
  \label{tab:rel16}
\end{table}

The 3rd Generation Partnership Project published a new release working on the upcoming 5G standard~\cite{3gpp_rel16}. Requirements has been defined and multiple use cases have been formed out of a set of these requirements. these use-cases define scenarios for 5G applications. A close fit to the described prototype is found in the use case 5.3.6 `Factories of the Future`.Two aspects of the this use case are of special interest for this prototype. These are shown in table \ref{tab:rel16} as an excerpt from release 16.

 The first aspect of the use case is defined as   
`Mobile control panels with safety functions; bi-directional communication` whereas aspect 2 expands on this with `... cyclic interaction; not more than two concurrent communication services of this type in the same service area.`~\cite{3gpp_rel16} 
Aspect 1 is used to compare the non-safety relevant communication in the prototype system. 
The examined area of the prototype is considerably smaller than the maximum size, roughly 20m x 20m. 
The message size is well in the range if only safety-relevant packets are considered. The majority of packets is non-safety-relevant, possessing larger size and frequencies than the ProfiSafe messages. 
The availability is subject to discussion. The most ambitious goal would be to reach the availability of a cable which is considered to be ~99,9999999\%. Both aspects are specified with lower availability. 
For safety certifications, this is of greatest importance as well as the potential risks of a missing package.
Aspect 2 considers a different scenario. The overall data rate of 5.97Mbit/s in the system is clearly above the specification of aspect 1 and therefore shows the need for a coexistence of aspects. The longer latency times pose no risk if only applied to the non-safety-relevant sensor systems, like RFID, etc. Please note, that a core safety functionality is still performed on the robot (laser range finder, etc.). 

Overall the prototype is well represented in the use case by combining aspects. This is comparable to other fields, like autonomous driving where different use cases are present in coexistence.

\section{Conclusion} 
In this work it is shown, that 5G can be applied in a real-world application industrial environment with heterogeneous data sources. It offers a solution to robotic communication with wireless emergency stop over Profisafe and data transmission to cloud systems. 
Additionally, a separation of perception and action from resources has been demonstrated which is especially useful for mobile robotics. 
The combination of multiple profiles to fulfill application specific requirements have been examined in this demonstrator. 

A flexible production allows to exploit concepts like Plug-n-Produce and predictive maintenance. Higher adaptivity may decrease down times caused by bottle necks in the production process. 
Various measurements have been performed to evaluate the proposed approaches in a real-world industrial environment. A comparison with an upcoming standard has been performed to evaluate if the prototype fits into the profiles. 

Overall, the capabilities of 5G communications show great potential for certain industrial applications as an alternative for wire-based transmission. 

\vfill\eject

\end{document}